\documentclass[12pt]{article}
\usepackage{geometry} % see geometry.pdf on how to lay out the page. There's lots.
\usepackage[T1]{fontenc}
\usepackage[utf8]{inputenc}
\usepackage[english]{babel}
\usepackage[english]{layout}
\usepackage{cite}

\usepackage{amsfonts,latexsym,amsthm,amssymb}
\usepackage{amsmath,amssymb,bbm,amsfonts}  % Better maths support & more symbols
\usepackage{bm}  % Define \bm{} to use bold math fonts
\usepackage{mathtools}
\usepackage[mathscr]{eucal}

\numberwithin{equation}{section}

\def\G{\Gamma}

\def\d{\delta}
\def\a{\alpha}
\def\b{\beta}
\def\e{\epsilon}
\def\f{\phi}

\def\l{\lambda}
\def\g{\gamma}
\def\e{\epsilon}
\def\m{\mu}
\def\n{\nu}

\def\s{\sigma}
\def\S{\Sigma}
\def\O{\Omega}
\def\o{\omega}
\def\th{\theta}
\def\pa{\partial}
\def\p{\pi}

\def\i{\iota}

\def\f{\phi}

\def\be{\begin{equation}}
\def\ee{\end{equation}}
\def\bea{\begin{eqnarray}}
\def\eea{\end{eqnarray}}
\def\R{{\bf R}}
\def\C{{\bf C}}

\def\A{{\cal A}}
\def\E{{\cal E}}
\def\sn{\textrm{sn}}
\def\cn{\textrm{cn}}
\def\dn{\textrm{dn}}
\def\pn{\textrm{pn}}
\def\qn{\textrm{qn}}
\def\pq{\textrm{pq}}
\def\sc{\textrm{sc}}
\def\dc{\textrm{dc}}
\def\nc{\textrm{nc}}
\def\ds{\textrm{ds}}
\def\cs{\textrm{cs}}
\def\ns{\textrm{ns}}
\def\nd{\textrm{nd}}
\def\cd{\textrm{cd}}
\def\sd{\textrm{sd}}
\def\arc{\textrm{arc}}
\def\min{\textrm{min}}

\def\ha{\frac{1}{2}}

\begin{document}

\title{\textbf{Self-dual cosmological Bianchi IX and VIII metrics}}

\author{ 
\textsf{A. ~Mikovi\'c}
\thanks{E-mail address: amikovic@ulusofona.pt} 
\textsf{~ and ~N. ~Manojlovi\'c}
\thanks{E-mail address: nmanoj@ualg.pt}\\
\\
\normalsize{\textit{$^{\ast \dag}$Grupo de F\'{\i}sica Matem\'atica da Universidade de Lisboa}}\\
\normalsize{\textit{Departamento de Matem\'atica, Faculdade de Ci\^encias}}\\
\normalsize{\textit{Campo Grande, Edif\'{\i}cio C6, PT-1749-016 Lisboa, Portugal}}\\
\\
\normalsize{\textit{$^{\ast}$Departamento de Matem\'atica e COPELABS}} \\
\normalsize{\textit{Universidade Lus\'ofona}}\\
\normalsize{\textit{Av. do Campo Grande, 376, 1749-024 Lisboa, Portugal}}\\
\\
\normalsize{\textit{$^{\dag}$Departamento de Matem\'atica, Faculdade de Ci\^encias e Tecnologia}}\\
\normalsize{\textit{Universidade do Algarve, Campus de Gambelas}}\\ 
\normalsize{\textit{PT-8005-139 Faro, Portugal}}\\
\\ 
}

\date{}

\maketitle
\thispagestyle{empty}
\begin{abstract}
We show that self-dual Bianchi IX and VIII cosmological models are described by the Nahm dynamical system for an appropriate type of matrices. We construct the general solutions in the case of the diagonal reductions of the corresponding Nahm equations and give the explicit expressions for the corresponding self-dual metrics in the Euclidean and in the Minkowski signature cases.
\end{abstract}

\clearpage
\newpage

\section{Introduction}

Imposing self-duality on the solutions of the Einstein equations is a way to generate new solutions, since one often ends up with an integrable system of differential equations \cite{DMW,Nea}. However, one can use this method only for the eucledean General Relativity (GR) or for the complex GR. Still, the eucledean or complex GR solutions are usefull, since they play a role in quantum gravity \cite{GP,HH}.

Self-dual Bianchi metrics have been mainly studied in the context of  self-duality of the Weyl tensor and spherical symmetry, see \cite{sdwb1,sdwb2, PP90, Tod91, Tod94, Hitchin95, Hitchin98, BK}. Self-dual spherically symmetric Bianchi metrics were studied in the context of self-duality of the Riemann tensor in \cite{I}. In this paper we will analyze the case of self-dual cosmological Bianchi metrics and a good framework for such a study is the Ashtekar formalism for self-dual metrics \cite{AJS}, as well as the Ashtekar formalism for the Bianchi cosmological spacetimes \cite{MM}.

Bianchi metric reductions of the Einstein equations give dynamical systems from classical mechanics \cite{MM02, MM03}, and imposing the Riemann tensor self-duality is expected to give integrable dynamical systems. This is reasonable to espect, because the spherically symmetric reduction of a self-dual metric leads to the Lagrange or the Halphen system of ordinary differential equations (ODE), see \cite{I, HMK00, HMKa00}. We will show that in the case of Bianchi IX and Bianchi VIII cosmological spacetimes one obtains the Nahm systems of ODE for the matrices form the Lie algebras $so(3)$ and $so(2,1)$, respectively. Also, in the case of the complex GR, the self-dual Bianchi IX and VIII reductions give the Nahm system for matrices from complex $so(3)$ and $so(2,1)$ Lie algebras. In the case of other Bianchi cosmological spacetimes, the self-dualty restriction gives linear systems of ODE, so that self-dual Bianchi IX and VIII cosmological models are the interesting cases as far as the integrability is concerned.

In section 2 we review the self-dual metrics in the Ashtekar formulation of GR. In section 3 we review the cosmological Bianchi models in the Ashtekar formulation  and show that in the Bianchi IX case the dynamical equations are given by the 3-dimensional real (or 2-dimensional complex in the Minkowski case) Nahm system of differential equations. In section 4 we analyze the integrability of the real and the complex Nahm system, and solve the diagonally reduced system, which is the Lagrange dynamical system. In section 5 we study the case of self-dual Bianchi VIII model, and solve the corresponding diagonally reduced Nahm system. In section 6 we construct the Bianchi IX and VIII self-dual metrics in the diagonally reduced cases. In section 7 we present our conclussions.

\section{Self-dual metrics in the Ashtekar formulation}

Let $M = \S \times \bf R$ be a 4-manifold where $\S$ is a 3-manifold. Let $g$ be a metric on $M$ and let $h$ be an induced metric on $\S$. These two metrics are related by
\be ds^2 = g_{\m\n}dx^\m dx^\n = \left(\xi \,N^2 + h_{ij}n^i n^j \right) dt^2 + 2dt dx^i h_{ij} n^j + h_{ij} dx^i dx^j \,,\label{4dcm}\ee
where $N$ is the laps, $n^i$ are the components of the shift vector and $\xi = 1$ in the Euclidean case while $\xi = -1$ in the Minkowski case. 

By using the metric (\ref{4dcm}) one can obtain the canonical formulation of the Einstein-Hilbert action, i.e. the Arnowitt-Deser-Misner formulation \cite{adm}, so that
\be \sqrt{|\det g|}\,R(g) \cong \p^{ij}\dot{h}_{ij} - n^i {\cal C}_i (\p,h) - N {\cal C}_0 (\p,h)\,,\label{adm}\ee
where $R(g)$ is the scalar curvature, $\cong$ is up to a surface term (when $\S$ is non-compact), $\p^{ij}$ are the canonically conjugate momenta for the 3-metric components $h_{ij}$ and $\dot X = dX/dt$. The constraints ${\cal C}_i$ and ${\cal C}_0$ are given by
\bea
{\cal C}_i &=& \nabla_j \, \p^{j}_{\,\,i} \\
{\cal C}_0 &=& {1\over\sqrt{\det h}} \left( \ha \p^2 - \p^{ij}\p_{ij} \right) + \sqrt{\det h} \, R(h) \,,
\eea
where $\p = h_{ij}\,\p^{ij}$.

One can change the canonical variables $(\p_{ij} , h^{ij})$ to $(p^i_\a , e_i^\a )$ canonical variables, where $e_i^\a$ are the triads, so that $h_{ij} = e_i^\a e_{j\a}$. Furthermore, one can pass to $(\tilde p_i^\a , \tilde e^i_\a )$ canonical variables, where $\tilde e^i_\a$ are densitized inverse triads given by 
\be \tilde e^i_\a = \sqrt{\det h}\, e^i_\a \,.\ee
Consequently
\be \sqrt{|\det g|}\,R(g) \cong \tilde p_{i}^\a{d{\tilde e}^{i}_\a \over dt} -  N^\a \tilde C_\a (\tilde p, \tilde e) -n^i \tilde C_i (\tilde p,\tilde e) - N \tilde C_0 (\tilde p,\tilde e)\,, \label{tadm}\ee
where $\tilde C_\m (\tilde p , \tilde e) = C_\m (p,e) = {\cal C}_\m (\p, h)$, $\m= 0$ or $\m=i$, and 
\be \tilde C_\a = \e_{\a\b\g}\,\tilde e^{i\b} \tilde p_i^\g \,.\ee
Here $\e_{\a\b\g}$ is the totally antisymmetric 3-dimensional symbol.

The Ashtekar variables are given by the canonical transformation 
$$(\tilde p_i^\a , \tilde e^i_\a )\to (A_i^\a , E^i_\a )$$ 
such that
\be A_i^\a = \G_i^\a (\tilde e) + z \tilde p_i^\a \,,\quad E^i_\a = \tilde e^i_\a \,,\label{acv}\ee
where $1 - \xi z^2 = 0$, $\G_i^\a(\tilde e) = \o_i^\a(e)$ and $\o_i^\a(e)$ is a spin connection on $\S$, whose dependence on the triads is given by the vanishing torsion equations
$$ T^\a = de^\a + \e^{\a\b\g}\,\o_\b \wedge e_\g = 0 \,. $$
The one-forms $A^\a$ are real in the eucledean gravity case ($z=\pm 1$), while they are complex in the Minkowski case ($z=\pm\,$i), and they are known as the Ashtekar connections \cite{A}.

By using (\ref{acv}), one can show that (\ref{tadm}) becomes
\be \sqrt{|\det g|}\,R(g) \cong -\xi z E^i_\a\dot A_i^\a - N^\a G_\a - n^i G_i - \tilde N G_0 \,,\ee
where $\tilde N = N/\sqrt{\det h}$,
$$ G_\a = D_i E^i_\a \,,\quad G_i = F_{ij}^\a \,E^j_\a \,,\quad G_0 = \e_{\a\b\g}\,F^\a_{ij} \,E^{i\b} E^{j\g} \,,$$
$D_i X = \pa_i X + [A_i,X]$ and $F = dA + [A,A]$ is the curvature 2-form for a real $SO(3)$ or $SU(2)$ connection $A$ in the Euclidean case, while in the Minkowski case we have a complex $SO(3)$ or $SU(2)$ connection $A$.

In the Euclidean gravity case the self-dual (SD) metrics are defined as
\be R_{ab}^* = R_{ab} \,,\ee 
where $R^{ab}$ is the curvature 2-form for the torsion-free spin connection $\o^{ab}$ on $M$ and 
$$R_{ab}^* = \e_{ab}^{\,\,\,\,\, cd}\, R_{cd} \,,$$ 
where $\e_{abcd}$ is the 4-dimensional totally antisymmetric symbol. The self-duality of the curvature is equivalent to the self-duality of the connection
\be \o_{ab}^*  = \o_{ab}\,,\label{sdm}\ee 
see \cite{EGH}. 

In the Minkowski gravity case the definition of self-duality as $X^* = X$ has to be modified, because $(X^*)^* = - X$. One can then define self-duality conditions as $X^* = \pm\,\textrm{i}\,X$, so that
\be  R_{ab}^* = \pm\,\textrm{i}\, R_{ab} \,\Leftrightarrow\,  \o_{ab}^* =\pm\,\textrm{i}\, \o_{ab}\,.\label{msd}\ee
The self-duality conditions (\ref{msd}) can be realized if we use complex metrics, which is also reflected by the fact that the Ashtekar connection is complex in the Minkowski signature case.

In the Euclidean case we can choose the gauge 
\be N^\a = 0\,,\quad n^i = 0 \,,\quad \tilde N = 1 \,,\ee
so that the 4-metric is given by
\be ds^2 = N^2 dt^2  + h_{ij} dx^i dx^j \,.\label{em}\ee
The Einstein equations are then given by
\bea 
\dot A_i^\a   &=& \e^\a_{\,\,\b\g} \, F_{ij}^\b E^{j\g} \,,\\
\dot E^i_\a   &=& \e_{\a\b\g} E^{j\b} D_j E^{i\g} \,,
\eea
plus the Gauss and the 3-diffeomorphism constraints
\be D_i E^i_\a = 0\,,\quad F_{ij}^\a E^j_\a = 0 \,. \ee

If we impose $A_i^\a = 0$, then $F_{ij}^\a =0$, which corresponds to the vanishing of the (anti) self-dual piece of the Riemann tensor, see \cite{B} for the Minkowski case. In the Euclidean case, $(\o_{ab}^*)^* = \o_{ab}$, so that
$$ \o_{ab} = \frac{1}{2}( \o_{ab} + \o^*_{ab} ) + \frac{1}{2}( \o_{ab} - \o^*_{ab} ) = \o^+_{ab} + \o^-_{ab} \,,$$
where $(\o_{ab}^\pm )^* = \pm\, \o_{ab}^\pm $ are the self-dual and the anti self-dual piece of the spin connection. Then $\o_{ab}^* = \o_{ab}$ corresponds to $R_{ab}^* = R_{ab}$, which corresponds to $R^-_{ab} = 0$ or $\o^-_{ab} = 0$. Hence the vanishing of the anti self-dual piece of the spin connection is equivalent to vanishing of the Ashtekar connection, since
$$\o^{-ab}_\m = 0 \Leftrightarrow \o^{-\a\b}_i = \frac{1}{2}\left( \o^{\a\b}_i - (\o^*)^{\a\b}_i \right)=0 \Leftrightarrow A_i^\a = \G_i^\a + z \tilde p_i^\a =0\,.$$

Hence the SD gravity equations are given by
\be \dot{E_\a} = \e_{\a\b\g}[E^\b , E^\g ] \,,\quad \pa_i E^i_\a = 0 \,,\label{sdea}\ee
where $E_\a = E^i_\a \,\pa_i$, which was the main result of  \cite{AJS}. The SD metric is given by (\ref{em}), where
\be N = \sqrt{\det h}\,,\quad h_{ij} = e_i^\a e_{j\a} \,, \quad e^i_\a = {E^i_\a \over \sqrt{\det h}}\,, \quad \det h = \det(E^i_\a)\,. \label{sdma}\ee

Note that the anti self-dual (ASD) metric equations are given by
$$\dot{E_\a} = -\, \e_{\a\b\g}[E^\b , E^\g ]\,,$$
while the other equations are the same as in the SD case.

In the Minkowski case the SD/ASD equations are given by
\be \dot{E_\a} = \pm\,\textrm{i}\, \e_{\a\b\g}[E^\b , E^\g ]\,,\ee
while the other equations are the same as in the Eucledean case except the expression for the 4-metric, which is given by
\be ds^2 = -N^2 dt^2 + h_{ij}dx^i dx^j \,.\ee

\section{Self-dual Bianchi cosmological models}

\noindent Bianchi cosmological spacetimes have topology $\S \times\R$ and globally defined 1-forms $\chi^I$ on $\S$ such that
$$ d\chi^I + C^{\quad I}_{JK} \chi^J \wedge \chi^K = 0 \,,$$
where $C^{\quad I}_{JK}$ are the structure constants of a 3-dimensional Lie algebra \cite{LL}. Bianchi showed that
$$ C^{\quad I}_{JK} = \e_{JKL}\, S^{LI} + \d^I_{[J} v_{K]} \,,$$
where $S$ is a diagonal matrix whose values can be $0$ or $\pm 1$ and $v_K =(v,0,0)$ \cite{LL}.

The inverse $\chi^I_i$ vector fields $L_I = L_I^i \pa_i$ satisfy the Lie algebra \cite{LL}
\be [L_I, L_J ] = C_{IJ}^{\quad K} \, L_K \,.\label{bla}\ee

In the Ashtekar formulation for GR, one can write
$$ A_i^\a (x,t) = \A_I^\a (t) \,\chi^I_i (x)\,,\quad E^i_\a (x,t) = \E^I_\a (t) \,L^i_I (x) \,,$$
see \cite{MM}, so that the self-dual equations of motion (EOM) (\ref{sdea}) become
\be \dot{\E^I_\a} = \e_\a^{\,\,\,\,\b\g} \,\E_\b^J \,\E_\g^K \, C_{JK}^{\quad I} \,,\quad \E^I_\a \, v_I = 0 \,,\label{sdb}\ee
and $\A =0$.

For the class A Bianchi models $v = 0$, so that there is only one equation, while for the class B, $v \ne 0$, so that a non-trivial solution ($\E \ne 0$) requires $\det \E = 0$.  

The dynamical equation from (\ref{sdb}) can be rewritten in the case of the Bianchi IX model ($S = diag(1,1,1)$, $v = 0$) as the Nahm equation
\be \dot V_\a = \frac{1}{2}\,\e_{\a}^{\,\,\,\,\b\g}\, [V_\b , V_\g ] \,,\label{ne}\ee
where $V_\a = 2\E_\a = 2\E_\a^I \, T_I$ and $T_I$ are $n\times n$ real matricies which generate the Lie algebra (\ref{bla}).

For $C_{IJK} = \e_{IJK}$ we can take $(T_I)_{JK} = -\,\e_{IJK}$, so that $n=3$.

We can also take $T_I = -\frac{\textrm{i}}{2}\,\s_I$, where $\s_I$ are Pauli matrices, so that $C_{IJK} = \e_{IJK}$ and $n=2$. This choice requires complex Nahm matrices, but if we restrict the Lie algebra coefficients to real numbers, we can still get real metric components. If we consider complex metrics, we can take the same Lie algebra generators for the $n=2$ and $n=3$ case and allow the coefficients to take complex values. In this case
\be \dot \E_\a = \textrm{i}\,\e_{\a}^{\,\,\,\,\b\g}\, [\E_\b , \E_\g ] \,,\label{cne}\ee
and $V_\a = 2\textrm{i}\,\E_\a$, then obeys the Nahm equation (\ref{ne}) for complex matricies $V_\a$.

As far as the integrability is concerned, the only non-trivial SD Bianchi models are Bianchi IX and VIII, since the other Bianchi models give linear systems of ODE.

\section{Self-dual Bianchi IX model}

The Nahm equations (\ref{ne}) can be written as
\be \dot V_1 = [V_2 , V_3 ] \,, \quad \dot V_2 = [V_3 , V_1 ] \,, \quad \dot V_3 = [V_1 , V_2 ] \,. \label{neom}\ee

The corresponding Lax pair is given by
\be L = A_+ + \l A_3 + \l^2 A_- \,,\quad M = \frac{1}{2}\,A_3 + \l A_- \,,\label{nlp}\ee
where
$$ A_\pm = V_1 \pm\textrm{i} V_2 \,,\quad A_3 = 2\textrm{i}\,V_3 \,, $$
so that the EOM (\ref{neom}) are equivalent to
$$ \dot L = [M, L] \,.$$

The integrals of motion $I_{mn}(V)$ can be determined from
\be tr(L^n ) = \sum_m I_{mn}(V) \l^m \,,\label{liom}\ee
where $n = 1,2,3,...$. We need to find 8 independent integrals of motion from the equation (\ref{liom}) in order to construct the general solution. However, due to the identity
$$ L^3 = \frac{1}{2}\,\textrm{tr}(L^2) L \,,$$
the only independent integrals of motion which are generated by tr$(L^n)$ are the 5 integrals coming from tr$(L^2)$,
and these are
$$I_1 = a^2 - b^2 \,,\quad I_2 = a^2 + b^2 - 2c^2 \,,\quad I_3 = \vec a \cdot \vec b \,, \quad I_4 = \vec b \cdot \vec c \,, \quad I_5 = \vec a \cdot \vec c \,,$$
where $\vec a = (u_1,v_1,w_1)$, $\vec b = (u_2,v_2,w_2)$, $\vec c = (u_3,v_3,w_3)$, $a^2 =(\vec a)^2$ and $b^2 =(\vec b)^2$. 

The components of these vectors are related to the elements of the 3-dimensional matrices $V_\a$ as
$$V_\a  = \left(
\begin{array}{ccc} 
0 & -w_\a & v_\a \\ 
w_\a & 0 & -u_\a \\
-v_\a & u_\a & 0 
\end{array}
\right) \,, $$
where $V_\a = u_\a \,T_1 + v_\a \,T_2 + w_\a \,T_3$. By choosing $u,v,w \in \R$ we get the Euclidean gravity case, while $u,v,w \in \C$ gives the Minkowski complex gravity case. 

In the 2-dimensional case, we have
$$V_\a  = \frac{1}{2}\left(
\begin{array}{cc} 
 -\textrm{i}w_\a & v_\a -\textrm{i}u_\a \\ 
v_\a + \textrm{i}u_\a  & \textrm{i}w_\a 
\end{array}
\right) \,, $$
so that even when  $u,v,w \in \R$ one obtains complex matrices. In order to avoid confusion, we will work with the 3-dimensional representation.

In terms of the vectors $\vec a$, $\vec b$ and $\vec c$ the Nahm system takes the form
\be \dot{\vec a} = \vec b \times \vec c \,,\quad \dot{\vec b} = \vec c \times \vec a \,,\quad\dot{\vec c} = \vec a \times \vec b \,.\label{vne}\ee

Although we cannot find a general solution of the Nahm system, we can find a general solution for a reduced Nahm system given by 
\be V_1 = x T_1 \,,\quad V_2 = y T_2 \,,\quad V_3 = z T_3 \,.\label{lr}\ee
In this case we obtain the Lagrange system
\be \dot x = y \, z  \,, \quad \dot y = x \, z  \,, \quad \dot z = x \, y \,.\label{rb9}\ee

The Lagrange system has quadratic integrals of motion
$$I = \a x^2 + \b y^2 + \g z^2 \,, \quad \a+\b+\g = 0 \,,$$
so that there are 2 independent quadratic integrals of motion.
By taking $C_1 = x^2 - z^2$ and $C_2 = y^2 - z^2$ we obtain $ x_i = \pm\sqrt{C_i + z^2}$, so that
$$ \dot z = \pm\sqrt{(C_1 + z^2 )(C_2 + z^2 )} \,.$$

Hence
\be t + C_3 = \int {dz \over \sqrt{(C_1 + z^2 )(C_2 + z^2 )}} \,,\label{zim}\ee
and a real solution is obtained in the following cases 
\be C_1 > 0 \,,\quad C_2 > 0 \,,\ee 
\be C_1 > 0 \,,\quad C_2 < 0\,,\quad z^2 > -C_2\,,\ee
\be C_1 < 0 \,,\quad C_2 > 0\,,\quad z^2 > -C_1\,,\ee
\be C_1 < 0 \,,\quad C_2 < 0\,,\quad z^2 > \textrm{max}\,\{-C_1 , -C_2 \}\,. \ee

In all these cases the solutions can be written as Jacobi elliptic functions \cite{WW, Ak, AS}. Let
$$ \pq (u,k) = {\pn(u,k) \over \qn (u,k)} \,,$$
where 
$$ \pn(u,k)\ne \qn(u,k)\,\in \{\sn(u,k), \cn(u,k), \dn(u,k) \}\,,$$ 
are the 3 basic Jacobi elliptic functions ($u\in\R$, $0 < k^2 < 1$, see \cite{AS}). 

Let $C_1 = \a^2$, $C_2 = \b^2$, $\a < \b$ and $u=\b(t+C_3)$. Then (\ref{zim}) becomes
$$ u = \int_0^{z/\a} (1+t^2)^{-1/2}[1+(k')^2 t^2]^{-1/2}\, dt = \arc\sc (z/\a , k)\,,$$
where $k'=\sqrt{1-k^2}= \a/\b$. Consequently
\be z = \a\, \sc(u,k)\,,\quad x = \b\,\dc(u,k)\,,\quad y = \a\, \nc(u,k)\,.\label{ab9}\ee

If $\a \ge \b$ then $u=\a(t+t_0)$, $k'=\b /\a$, and $x,y$ and $z$ are given by (\ref{ab9}) with $(\a,\b)\to(\b,\a)$.

In the case $C_1 = -\a^2$, $C_2 = \b^2$, we have $u = \g(t+t_0)$ and 
$$ u = \int_{z/\g}^{+\infty} (t^2 - k'^2)^{-1/2}( t^2 + k^2)^{-1/2}\, dt = \arc\ds \left(z/\g,k \right)\,,$$
where $\g = \sqrt{\a^2 + \b^2}$ and $k = \frac{\a}{\g}$. Then
\be z =  -\g\, \ds (u,k)\,,\,\, x = -\g\,\cs (u,k)\,,\,\, y = -\g \,  \ns (u,k)\,.\label{bb9}\ee

In the last case we have $C_1 = -\a^2$, $C_2 = -\b^2$ and for $\b < \a$ we obtain $u=\a(t+t_0)$ and
$$ u = \int_{z/\a}^{+\infty} (t^2 - 1)^{-1/2}( t^2 - k^2)^{-1/2}\, dt = \arc\dc (z/\a, k) \,,$$
where $k = \b /\a$. Consequently
\be z = \a\,\dc (u,k)\,,\quad x = \g\,\sc (u,k)\,,\quad y = \g\,\nc (u,k)\,,\label{cb9}\ee
where $\g = \sqrt{\a^2 -\b^2}$.

When $\a < \b$, the solution is given by (\ref{cb9}) with $(\a,\b)\to (\b,\a)$, so that $u=\b(t+t_0)$ and $k =\a /\b$.

In the complex case one can use  the solution of (\ref{rb9}) given in \cite{T}, which can be written as 
\be x = \sqrt{I_1}\,\sn ( u, k)\,,\quad y = \sqrt{-I_1}\,\cn (u , k)\,,\quad
     z = -\sqrt{-I_2}\,\dn (u , k) \,, \label{tsol} \ee
where 
$$ k  = {\sqrt{I_2} - \sqrt{I_2 - I_1} \over \sqrt{I_1}} \,,\quad u = \sqrt{I_2}\, (t + t_0 ) \,,$$
and
$$I_1 = x^2 - y^2 \,,\quad I_2 = x^2 - z^2 \,.$$

Note that a complex solution can be also constructed by taking one of the real solutions (\ref{ab9}), (\ref{bb9}) or (\ref{cb9}) and substituting $\a$ and $\b$ by $\pm\sqrt{C_1}$ and $\pm\sqrt{C_2}$. For example, the real solution (\ref{ab9}) gives a complex solution
\be x = \sqrt{C_2}\,\dc(u,k)\,,\quad y = \sqrt{C_1}\, \nc(u,k)\,,\quad z = \sqrt{C_1}\, \sc(u,k)\,,\label{cab9}\ee
where
$$ k = \sqrt{{C_2 - C_1 \over C_2}} \,,\quad u = \sqrt{C_2} (t + t_0 ) \,,$$
and there is no sign restriction on $C_1$ and $C_2$.

\section{Self-dual Bianchi VIII model}

In the Bianchi VIII case $S=diag(1,1,-1)$ and $v = 0$, so that we have the system
\be  \dot{\vec a} = \vec b \times \vec c \,,\quad \dot{\vec b} = \vec c \times \vec a \,,\quad\dot{\vec c} = -\vec a \times \vec b \,.\label{b8eom}\ee

The Lax pair is given by (\ref{nlp}), where now the matrices $V_\a$ belong to the $so(2,1)$ Lie algebra, so that
$$V_\a  = \left(
\begin{array}{ccc} 
0 & -w_\a & v_\a \\ 
w_\a & 0 & u_\a \\
v_\a & u_\a & 0 
\end{array}
\right) \,. $$
The quadratic integrals of motion are then given by
\be I_1 = a^2 - b^2 \,,\quad I_2 =  a^2 + b^2 +2 c^2 \,,\quad I_3 = \vec a \cdot \vec b \,, \quad I_4 = \vec b \cdot \vec c \,, \quad I_5 = \vec a \cdot \vec c \,.\label{b8i}\ee

Note that $I_1$ and $I_2$ are two independent integrals of motion coming from a set of quadratic integrals of motion
$$I = \a\, a^2 + \b \,b^2 + \g\, c^2 \,, \quad \a+\b - \g = 0 \,.$$
Hence we can also chose the pair $I_1' = a^2 + c^2$ and $I_2' = b^2 + c^2$ instead of the $(I_1,I_2)$ pair.

As in the Bianchi IX case, the integrals (\ref{b8i}) are not sufficient to solve the system (\ref{b8eom}), and we can make the reduction
$$ \vec a = (x,0,0) \,,\quad \vec b = (0,y,0) \,,\quad \vec c = (0,0,z) \,.$$
In this case we get a modified Lagrange system
\be \dot x = y \, z  \,, \quad \dot y = x \, z  \,, \quad \dot z = - x \, y \,.\label{rb8}\ee
By using the integrals of motion $a^2 + b^2$ and $b^2 + c^2$, we obtain
\be C_1 = x^2 + z^2 \,,\quad C_2 = y^2 + z^2 \,,\ee
so that
\be\dot z = -\sqrt{(C_1 - z^2)(C_2 - z^2)}\,.\ee

Hence
\be t + t_0 = - \int {dz \over \sqrt{(C_1 - z^2 )(C_2 - z^2 )}} \,,\label{b8rs}\ee
and real solutions are obtained for 
\be C_1 > 0 \,,\quad C_2 > 0 \,,\quad z^2 < \min\{C_1, C_2\}  \,.\ee

Let $C_1 = \a^2$, $C_2 = \b^2$, $\a<\b$ and $u=\b(t+t_0)$ then (\ref{b8rs}) becomes
$$ -u = \int_0^{z/\a} (1-t^2)^{-1/2}(1- k^2 t^2)^{-1/2}\, dt = \arc\sn (z/\a , k)\,,$$
where $k=\a /\b$. Hence
\be z = \a\, \sn (-u,k) \,,\quad x= \a\,\cn(-u,k)\,,\quad y= \b\,\dn(-u,k) \,. \label{b8a}\ee

There is an alternative form of the reduced Bianchi VIII solution: if we take the integrals
$$ x^2 - y^2 = C'_1 \,,\quad y^2 + z^2 = C_2 \,,$$
then we obtain
$$\dot y = \sqrt{(C'_1 + y^2)(C_2 - y^2)}\,.$$

The real solutions are obtained for $C'_1 > 0$ and $C_2 > 0$ or for $C'_1 < 0$ and $C_2 > 0$. In the first case let $C_1' = \a^2$ and $C_2 = \b^2$, then
$$ u = \int_{y/\g}^1 (1-t^2)^{-1/2}(k'^2 + k^2 t^2)^{-1/2}\, dt = \arc\cn (y/\b , k)\,,$$
where $\g=\sqrt{\a^2 + \b^2}$, $u=\g (t+t_0)$, $k=\b /\g$. Consequently
\be y = \b\,\cn\,(u,k) \,, \quad x=\g\,\dn\, (u,k) \,,\quad z= -\b\,\sn\,(u,k)\,.\label{b8b} \ee

In the second case let $C_1' = -\a^2$, $C_2 = \b^2$ and $\b > \a$, then
$$ u = \int_1^{y/\a} (t^2 -1)^{-1/2}(1 -  k'^2 t^2)^{-1/2}\, dt = \arc\nd (y/\a , k)\,,$$
where $u=\b (t+t_0)$, $k'=\a /\b$. Consequently
\be y = \a\,\nd \,(u,k) \,, \quad x= \a'\,\sd\, (u,k) \,,\quad z= \b'\,\cd\,(u,k) \,, \label{b8c}\ee
where $\a' = \a\sqrt{1 - (\a/\b)^2}$ and $\b' = \b\sqrt{1 -(\a/\b)^2}$.

In the complex metric case one can take any of the real solutions (\ref{b8a}), (\ref{b8b}) or (\ref{b8c}), and put $\a=\sqrt{C_1}$ or $\a=\pm\sqrt{C'_1}$, $\b=\sqrt{C_2}$ whith no restriction on the signs of the constants. For example, the real solution (\ref{b8a}) gives a complex solution
\be x= \sqrt{C_1}\,\cn(-u,k)\,,\quad y= \sqrt{C_2}\,\dn(-u,k) \,,\quad z = \sqrt{C_1}\, \sn (-u,k) \,, \label{cb8a}\ee
where
$$ k = \sqrt{{C_1 \over C_2}} \,,\quad u = \sqrt{C_2} (t + t_0 ) \,,$$
and there is no sign restriction on $C_1$ and $C_2$.

\section{Cosmological self-dual Bianchi metrics}

By using the definitions (\ref{em}), (\ref{sdma}) and (\ref{lr}) one obtains that the metric in the reduced (Lagrange) case is
\be ds^2 = \O_1\O_2\O_3 (\det L )\left( dt^2 + {\chi_1^2 \over\O_1^2}+{\chi_2^2 \over\O_2^2}+{\chi_3^2 \over\O_3^2}\right)\,,\label{sdbm}\ee
where $\O_i = x_i(t)/2$ and $\det L =\det(L^i_\a)$. 

In the Bianchi IX case we have
\bea L_1 &=& \sin\psi \,\pa_\th  - {\cos\psi\over\sin\th}\,\pa_\f + \cot\th \cos\psi\,\pa_\psi \,,\nonumber\\
L_2 &=& \cos\psi \,\pa_\th  + {\sin\psi\over\sin\th}\,\pa_\f - \cot\th \sin\psi\,\pa_\psi \,, \\
L_3 &=& \pa_\psi \,, \nonumber\eea
where $\th,\f,\psi$ are the Euler angles\footnote{We use the parametrization $R(\f,\th,\psi) = R_3(\psi)\,R_1(\th)\,R_3 (\f)$ for an $SO(3)$ group element, where $R_1$ and $R_3$ are rotations around the $x$ and the $z$ axis.}, while $x_i(t)$ are given by (\ref{ab9}),(\ref{bb9}) and (\ref{cb9}). 

Consequently
$$ \det(L^i_\a) = {1\over\sin\th} \,,$$
and
\bea \chi_1 &=& \sin\psi \,d\th  - \cos\psi\,\sin\th\,d\f \,,\nonumber\\
 \chi_2 &=& \cos\psi \,d\th  + \sin\psi \sin\th\,d\f \,,\label{b9f}\\
 \chi_3 &=&  \cos\th\,d\f + d\psi \,, \nonumber\eea
so that
\be ds^2 = {\O_1\O_2\O_3 \over\sin\th}\left( dt^2 + {\chi_1^2 \over\O_1^2}+{\chi_2^2 \over\O_2^2}+{\chi_3^2 \over\O_3^2}\right)\,.\label{sdb9m}\ee

In the Bianchi VIII case we have 
%the same formulas but $\f\to i\f$ and $\psi\to i\psi$ (verify) 
\bea L_1 &=& \sin\psi \,\pa_{\th'}  - {\cos\psi\over\sinh\th'}\,\pa_\f + \coth\th' \cos\psi\,\pa_\psi \,,\nonumber\\
L_2 &=& \cos\psi \,\pa_{\th'}  + {\sin\psi\over\sinh\th'}\,\pa_\f - \coth\th' \sin\psi\,\pa_\psi \,, \\
L_3 &=& \pa_\psi \,, \nonumber\eea
where the angles are defined by an $SO(2,1)$ group element parametrization $R(\f,\th,\psi) = R_3(\psi)\,\tilde R_1(\th')\, R_3 (\f)$ such that
$$ \tilde R_1 (\th') =\left(
\begin{array}{ccc} 
1 & 0 & 0\\
0 &\cosh\th' & \sinh\th'  \\ 
0 &\sinh\th' & \cosh\th' 
\end{array}
\right) \,,$$
and $\th'\in \bf R$.

Consequently
$$ \det(L^i_\a) = {1\over\sinh\th'} \,,$$
and
\bea \chi_1 &=& \sin\psi \,d\th'  - \cos\psi\,\sinh\th'\,d\f \,,\nonumber\\
 \chi_2 &=& \cos\psi \,d\th'  + \sin\psi \sinh\th'\,d\f  \,,\label{b8f}\\
 \chi_3 &=& \cosh\th'\,d\f + d\psi \,, \nonumber\eea
while the metric is given by (\ref{sdb9m}). The $\O_i(t)$ functions in the Bianchi VIII case are given by (\ref{b8a}).

In the case of a  Minkowski signature self-dual Bianchi IX complex metric, we have the same formulas for the $L$'s and the $\chi$'s as in the eucledean case, while $\O_i(t)$ are given by (\ref{tsol}) or by (\ref{cab9}) and
\be ds^2 = {\O_1\O_2\O_3 \over\sinh\th'}\left( - dt^2 + {\chi_1^2 \over\O_1^2}+{\chi_2^2 \over\O_2^2}+{\chi_3^2 \over\O_3^2}\right)\,.\label{cmsdb}\ee

In the case of a complex Bianchi VIII self-dual metric we have (\ref{cmsdb}) with $\O_i(t)$ given by (\ref{cb8a}).

\section{Conclusions}

The main results are the self-dual Bianchi IX and VII cosmological metrics given by the expression (\ref{sdb9m})  in the Euclidean case, and by the expression (\ref{cmsdb}) in the Minkowski case. The forms $\chi^I$ are given by (\ref{b9f}) in the Bianchi IX case and  by (\ref{b8f}) in the Bianchi VIII case. The functions $2\O_i(t)$ are given by (\ref{ab9}),(\ref{bb9}) and (\ref{cb9}) in the Bianchi IX case, while in the Bianchi VIII case these functions are given by (\ref{b8a}). In the Minkowski case,  the functions $2\O_i(t)$ are given by (\ref{cab9}) for the Bianchi IX case, while in the Bianchi VII case these functions are given by (\ref{cb8a}).

Note that we solved the reduced Nahm's equations (\ref{rb9}) and (\ref{rb8}), so that a natural next step would be to solve the complete set (\ref{vne}) or (\ref{b8eom}). This would require the knowledge of 8 independent integrals of motion, and the Lax method gives only 5. This means that one could solve a 6-variables reduction of the Nahm's equations. However, it is not clear how to implement a 6-variables reduction such that it is preserved by the time evolution. As far as solving the complete set of Nahm's equations is concerned, one would need to find additional 3 integrals of motion, and there are indications that these conserved quantities cannot be local functions.

Note that in \cite{I} it was considered a selfdual spherically-symmetric metric of the form
\be ds^2 = \O_1\O_2\O_3 \left( dr^2 + {\s_1^2 \over\O_1^2}+{\s_2^2 \over\O_2^2}+{\s_3^2 \over\O_3^2}\right)\,,\label{spsdm}\ee
where $\O_i = f_i(r)$, $r^2 = x^2 + y^2 + z^2 + t^2$,
$$ \s_i = \frac{1}{r^2}\,\eta_{i\m\n}x^\m dx^\n \,,$$
and $\eta_{i\m\n}$ are the t'Hooft coefficients.

The metric (\ref{spsdm}) has a similar structure as the SD Bianchi IX metric (\ref{sdbm}), but the variables are different. In \cite{I} it was also showed that the self-duality of the connection associated to the metric (\ref{spsdm}) gives the Lagrange system 
$$\O'_i = \O_j \O_k \,,\quad i\ne j\ne k \,.$$
%while the self-duality of the Reimann tensor gives the Halphen system
%$$\O'_i = \O_j \O_k - \O_i (\O_j + \O_k) \,.$$
Hence our Lagrange system solutions (\ref{ab9}),(\ref{bb9}) and (\ref{cb9}) can be used to construct self-dual metrics of the type (\ref{spsdm}), simply by replacing the variable $t$ in $x_i(t)$ with the variable $r$.

\end{document}